\documentclass[conference]{IEEEtran}
\ifCLASSINFOpdf
\else
\fi
\usepackage{times,amsmath,epsfig}
\usepackage{epstopdf}
\usepackage{algpseudocode}
\usepackage{algorithm}
\usepackage{mathtools}
\usepackage{balance}
\usepackage{color}
\usepackage{amsfonts}
\usepackage{bbm}
\DeclarePairedDelimiter{\abs}{\lvert}{\rvert}

\DeclareMathOperator*{\argmax}{arg\,max}

\usepackage{subcaption}
\usepackage{cleveref} 
\begin{document}
%
\title{Fast Grant Learning-Based Approach for Machine Type Communications with NOMA}


\author{\IEEEauthorblockN{Manal El Tanab and  Walaa~Hamouda}
\IEEEauthorblockA{ Department of Electrical and Computer Engineering, Concordia University,\\
 Montreal, Qc, Canada, H3G 1M8\\
Email: meltanab@encs.concordia.ca, 
hamouda@ece.concordia.ca}}


\maketitle
\thispagestyle{empty}
\pagestyle{empty} 

\begin{abstract}
In this paper, we propose a non-orthogonal multiple access (NOMA)-based communication framework that allows machine type devices (MTDs) to access the network while avoiding congestion. The proposed technique is a 2-step mechanism that first employs fast uplink grant to schedule the devices without sending a request to the base station (BS). Secondly, NOMA pairing is employed in a distributed manner to reduce signaling overhead. Due to the limited capability of information gathering at the BS in massive scenarios, learning techniques are best fit for such problems. Therefore, multi-arm bandit learning is adopted to schedule the fast grant MTDs. Then, constrained random NOMA pairing is proposed that assists in decoupling the two main challenges of fast uplink grant schemes namely, active set prediction and optimal scheduling. Using NOMA, we were able to significantly reduce the resource wastage due to prediction errors. Additionally, the results show that the proposed scheme can easily attain the impractical optimal OMA performance, in terms of the achievable rewards, at an affordable complexity.

\end{abstract}

\IEEEpeerreviewmaketitle

\section{Introduction}

Massive machine type communications (mMTC) and ultra-reliable low-latency communications (URLLC) are two main categories of machine-to-machine (M2M) networks that target massive and mission critical Internet of things (IoT) applications, respectively \cite{IoT_NOMA_MTC}. Mainly, M2M networks are characterized by small data packets, mostly in the uplink direction with heterogeneous quality of service (QoS) requirements. Unlike human-type communications that focuses on achieving high data rates, MTC have different requirements on connectivity, latency, and reliability.   
Congestion is a major challenge of MTC access that  leads to severe access delays and even service interruption \cite{Manal_ICC_scalable}. The legacy coordinated access technique for cellular networks was shown to be inefficient for MTC due to the large signaling overhead and high latency \cite{Rana_grantFree}. On the other hand, fully uncoordinated access, known as grant-free where the devices transmit directly on a random uplink resource without sending scheduling requests, reduces the signaling cost at the expense of collisions \cite{fast_grant_chalOpp}. Having much larger number of devices than the available resources is the likely case of MTC which reflects that collisions will degrade the performance. In \cite{3GPP_latencyReduction}, the 3GPP proposed fast uplink grant technique for latency reduction. In fast uplink grant, the base station (BS) allocates the uplink grants directly to the MTC devices (MTDs) without receiving any scheduling requests. Hence, both large signaling overhead and collisions are avoided. However, if an inactive MTD received an uplink grant, the resource is wasted. 


Fast uplink grant has two main challenges \cite{fast_grant_chalOpp}. First, the BS has to predict the set of active MTDs at each cycle. Second, the granted resources should go to the optimal MTDs according to each network requirements. This should be done with limited or no information of the devices’ QoS requirements or channel state information (CSI) which are very difficult to acquire for massive number of MTDs. Learning techniques are known to be very promising in solving problems in such environments. In particular, the authors in \cite{MAB_GLBCM} proposed a technique based on multi-armed bandit (MAB) theory which is essentially a class of reinforcement learning problems. Although this technique gives promising results, it has a major problem that is its performance is so dependent on the accuracy of the traffic prediction scheme. 
Hence, there is a need to decouple the two challenges of fast grant schemes or even reduce the effect of the predictor efficiency on the scheduler performance.

Recently, NOMA has received great attention as a promising enabling technique for beyond 5G wireless networks \cite{NOMA_5G_survey}, \cite{NOMA_status}. 
It allows multiple users to non-orthogonally share the same resource by multiplexing them either in power or code domain. 
From the information theoretic perspective, orthogonal multiple access (OMA) is strictly suboptimal in multi-user systems \cite{NOMA_theory}. Thus, using NOMA with MTC can boost the spectral efficiency while handling the massive connectivity and reducing latency. These promising gains come at the expense of a more complex receiver that is able to decode the superposed signals. For uplink NOMA, successive interference cancellation (SIC) decoder exists at the BS which is fortunately acceptable for low-budget MTDs with mostly-uplink traffic. 

In literature, variants of NOMA with MTC can be found.  In \cite{RA-based_NOMA_m2m}, a random access (RA)-based NOMA scheme for MTC was proposed. NOMA was also adopted with grant-free to avoid the large overhead of the RA process \cite{Rana_grantFree}. Moreover, a semi-grant free proposal was introduced to get advantages of both grant-based and grant free \cite{semi-grantFree_NOMA}. With fast uplink grant, there is a risk of wasting the scarce resources if an inactive MTD was scheduled. Although the literature has variants of NOMA with MTC, there is no work that combined NOMA with fast uplink grant. To the best of our knowledge this is the first work to study fast uplink grant for MTC with NOMA.




In this paper, we propose NOMA-based fast uplink grant for MTC to enhance the overall system performance and approach the decoupling of the predictor and scheduler performances. By allowing multiple MTDs to share the same resource, the resource wastage due to prediction errors can decrease. Obviously, this comes with the cost of extra signaling and complexity. However, starting with the simple 2-user NOMA can efficiently handle the tradeoff between system performance and complexity. For proper SIC decoding, certain level of distinctness between the received signals at the BS should be maintained \cite{Dynamic_ekram}. 
In literature, centralized pairing is commonly done by the BS to sort the users' channels and pair distinctive ones. However, for massive MTC, sending all devices’ CSI to the BS would result in expensive signaling. The goal of this proposal is to pair NOMA users in a distributed manner while enabling successful SIC decoding at the BS. 



The rest of the paper is organized as follows. In Section~\ref{sys-Mod}, the system model is presented. The problem formulation is in Section~\ref{prob formulation}. Section~\ref{proposed_scheme} explains the proposed scheme. In Section~\ref{optNOMA}, optimal NOMA is discussed. Simulation results are given in Section~\ref{sim}. Section\ref{conc} draws some conclusions.

\section{System Model}
\label{sys-Mod}
Consider a single cell network with $N$ MTDs are served by a single BS. Assume that the available bandwidth is divided into $M$ resource blocks (RBs), each of size $B$. At each time slot (i.e. cycle) the BS allocates the RBs to the available ``active'' MTDs using fast grant. The term ``active'' refers to an MTD with a packet ready for transmission within a predefined maximum tolerable delay. If a packet waits for access more than its maximum tolerable delay, it will be dropped out. A packet's access delay is the number of cycles elapsed from the moment it is ready for transmission until it is granted access. Due to the  heterogeneity of IoT applications, the packets have different QoS requirements that are unknown to the BS.

To model congestion, at the beginning, we consider a Beta-distributed activation of the MTDs as suggested by the 3GPP for overload situations \cite{3GPP-Beta}. Particularly, the $N$ MTDs are activated within a bounded activation time $T_A$ that is assumed to be divided into $I_A$ time slots. Considering a Beta distribution with parameters $(\alpha=3, \beta= 4)$, each MTD is activated at time $t$ with probability $A(t)$ as follows:
\begin{equation}
\label{A_calc}
A(t)= \frac{t^{\alpha-1} (T_A -t)^{\beta-1}}{T_A^{\alpha+\beta-1} \mathcal{B}(\alpha,\beta)},
\end{equation}
where $\mathcal{B}(\alpha,\beta)$ is the Beta function. 
This more realistic model is only adopted during the first activation of the devices instead of the uniform activation model adopted in \cite{MAB_GLBCM}. Then, at each cycle, a random set of inactive MTDs is selected for activation to have a dynamic activation process that is needed to build a history of each MTD for the learning algorithm.

For uplink NOMA, multiple MTDs transmit non-orthogonally to the same receiver (i.e. BS). At the BS, the signals are decoded using SIC. Hence, certain level of {\it distinction} has to be maintained. All channel gains are modeled as Rayleigh fading and assumed to be independent and identically distributed (i.i.d) across users and time. Moreover, both path loss and log-normal shadowing are considered. For 2-user NOMA, the received signal at the BS on the $m^{th}$ RB is:
\begin{equation}
\mathbf{y}_m= \sqrt{P_{s}}h_{sm}\mathbf{x}_{sm} + \sqrt{P_{w}}h_{wm}\mathbf{x}_{wm} + \mathbf{z}_m,
\end{equation}
where $P_{s}, h_{sm}, P_w, h_{wm}$ are the transmission power and channel gain for the strong and weak MTDs of the $m^{th}$ NOMA pair, corresponding to their signals $\mathbf{x}_{sm}, \mathbf{x}_{wm}$, respectively. $\mathbf{z}$ is an additive white Gaussian noise with zero mean and power spectral density $N_0$. We also assume that each MTD has a maximum power budget $P_t$.

The BS decodes the strong user's signal first, then cancels it and decodes the weak user's signal. Hence, only the strong user will suffer from interference. Based on that, the achievable rates of the NOMA pair can be expressed as:
\begin{eqnarray} 
r_s=& B\log_2 \left( 1+  \frac{P_s \gamma_s}{P_w\gamma_w+1}\right),\\
r_w=& B\log_2 \left( 1+ P_w \gamma_w\right),
\end{eqnarray}
where $\gamma=\frac{\abs{h}^2}{N_0 B} $ is the normalized channel gain. The subscripts $(.)_s,(.)_w$ denote the strong and weak users, respectively.

\section{Problem Formulation}
\label{prob formulation}
The heterogeneity of the QoS requirements of huge number of MTDs complicates the resource allocation process. In such environment, the optimal OMA scheduler aims to schedule $M$ MTDs at each time instant $t$ such that the utility is maximized under QoS constraints. This can be formulated as: 
\begin{eqnarray}
\label{U_opt}
\mathcal{S}(t) =\argmax_{\{i_1, \dots, i_M\} \in \mathcal{K}(t)}&\sum_{i=i_1}^{i_M} U_i(t)&\\
\text{s.t.} &r_i(t) \ge R_{i_{min}},& i= i_1, \dots, i_M\ \notag \\
&d_i(t) \le D_{i},& i= i_1, \dots, i_M\notag 
\end{eqnarray}
where $\mathcal{K},\mathcal{S}$ are the sets of active and scheduled MTDs, respectively. $U_{i}(t)$ is the utility of the scheduled MTD $i$.  
$R_{i_{min}}, D_{i} $ are the minimum rate and maximum delay constraints of MTD $i$, whereas $r_i(t), d_i(t)$ are the achieved rate and access delay for MTD $i$ at $t$. For massive MTC, delay requirements are commonly the most important. However, other QoS metrics can be imposed. Hence, the utility function $U$ is defined as a combination of different normalized QoS metrics as \cite{MAB_GLBCM}: 
\begin{equation}
\label{utility_fn}
U_i(t)= \delta_1 v_i(t) +\delta_2 r_i^n(t) + \delta_3 f(D_i(t)),
\end{equation}
where $\delta_1, \delta_2, \delta_3$ are weights for the importance of each metric such that their summation is $1$.
$v_i(t)$ is the value of information of the generated packet at time $t$ by MTD $i$ which assesses the importance of this packet in certain context. 
 The normalized value represents a percentage of importance, hence $v_i(t) \in [0,1]$. $r_i^n(t)$ is the normalized rate of MTD $i$ at $t$ that is obtained by dividing the achieved rate $r_i(t)$ by the maximum rate $R_{max}$ that could be achieved by the node having the best channel to the BS. In our setting, an arbitrary relatively-high value for $R_{max}$ was selected. For the maximum tolerable access delay $D_i(t)$, the normalization was obtained using a modified Gompertz function with parameters $a, b,c$ as follows:
\begin{equation}
\label{Gomp_delay}
f(D_i(t))= a-a e^{-b e^{-c D_i(t)}}.
\end{equation}

Solving the problem in \eqref{U_opt} requires the BS to gather a lot of information about the MTDs in the network. For instance, the BS needs to know the QoS requirements of each active MTD at each time instant, as well as the CSI for optimal throughput. However, gathering all this information for massive number of devices, usually with small-sized data packets, is highly inefficient. In the following, online learning combined with NOMA will be employed to solve the problem.


\section{Proposed Scheme}
\label{proposed_scheme}
The idea of the proposed scheme is to employ both fast grant and NOMA such that each granted uplink resource is shared by multiple MTDs. As proposed in \cite{MAB_GLBCM}, the MAB learning approach is used to enhance the performance of fast grant where the BS has no information about the MTDs' channels or QoS. It is worth mentioning that, a learning parameter in the 
 approach of \cite{MAB_GLBCM} is unknown to the BS. This was ignored in \cite{MAB_GLBCM}, but we take it into consideration as will be shown later. 

The proposed scheme is a 2-step as follows. First,  MAB is used such that the BS can select the scheduled devices for uplink grant \cite{MAB_GLBCM}. Second, the scheduled devices are considered as cluster heads (CHs) and seek pairing with other ``nearby'' devices that satisfy a power tolerance condition announced by the BS for successful SIC operation. 
After transmission, the BS receives QoS-based rewards from the scheduled MTDs. This method avoids the resource wastage inherited in fast uplink grant. The reason is that, if a device receives a grant and it has no ready packet for transmission, it can still seek pairing and forward the grant to other ``active'' MTDs that will respond to its pairing request. In the following, we show how the preceding traffic prediction step is abstracted and give details of both stages of the proposed scheme.

\subsection{Traffic Predictor Abstraction}
We assume that the BS employs a predictor with certain average prediction error $\bar{e}_p$. At each cycle, there are an actual active MTDs list and a BS prediction list that satisfies a per-cycle error $e_p \sim \mathcal{N}(\bar{e}_p, \sigma_e^2)$ truncated in [0,1]. 
We define $e_p$ as the total number of errors in prediction to the total number of MTDs in the network. Furthermore, the predictor outputs a probability of being active $P_a$ for each MTD.


\subsection{Fast Grant Scheduling using MAB}
Generally, in MAB problems there is a set of available arms where a decision maker selects an arm and observes the resulted reward aiming to maximizing  the cumulative reward. The distributions of the rewards of different arms are unknown to the decision maker. In our problem, the BS is the decision maker and the MTDs are the arms. According to \eqref{U_opt}, the reward of the $i^{th}$ MTD is defined as:
\begin{equation}
\label{Rwd_eqn}
\theta_i(t)= \mathbbm{1}[r_i(t) \ge R_{i_{min}}]  \mathbbm{1}[d_i(t) \le D_{i}  ] U_i(t),
\end{equation}
where $\mathbbm{1}(.)$ is an indicator function that gives $1$ if its argument holds and $0$ otherwise. Let us define the regret as the difference between the rewards of the best arm that could have been played and the selected arm. In terms of regret, the goal of the scheduler is to minimize the cumulative regret $\mathcal{R}$. Let $\theta_k (t)$ be the achieved reward of playing arm $k$ at time $t$, and $\theta^*(t)$ be the maximum reward that could have been achieved at time $t$, the regret up to time $T$ is as follows:
\begin{equation}
\label{old_regret}
\mathcal{R}(T)= \mathbb{E} \left[ \sum_{t=1}^{T} \theta^*(t) - \sum_{t=1}^{T} \theta_k (t)  \right].
\end{equation}

%

To maximize the cumulative reward, the upper-confidence bound (UCB) concept is used to solve the problem. UCB is well-known to achieve balance between exploitation and exploration. 
This well-known machine learning tradeoff requires balancing reward maximization based on exploiting the knowledge already acquired while attempting (i.e. exploring) new actions to further increase knowledge.
Since, the availability of the arms of MTC is probabilistic, sleeping MAB is more suitable. In sleeping MAB, at each time instant, only a subset of the arms is available. Given that the BS has a prediction algorithm, at each time slot $t$, it has the set of active MTDs $\mathcal{K}_t$ associated with certain probability of being active ${P_a}_i(t)$. 
 Then, the BS employs the UCB to play an arm $k(t)$ such that:
\begin{equation}
\label{Psi_eqn}
k(t)= \arg\!\max_{i \in \mathcal{K}_t}  {P_a}_i(t)  \left(\frac{z_i(t)}{n_i} + \sqrt{\frac{8 \log t^\prime}{n_i}}\right),
\end{equation}
where $z_i(t)$ is the sum of rewards of MTD $i$ up to time $t$, $n_i$ is the number of times MTD $i$ was selected and was active, and $t^\prime$ is the total number of times the selected MTD was active. Note that all the parameters in \eqref{Psi_eqn} are known to the BS except of $t^\prime$.  Using the traffic predictor, the estimate of $t^\prime$ is assumed to be the number of times each MTD was estimated active by the BS.
In this scheme, the BS selects the newly activated MTDs first before starting \eqref{Psi_eqn}. 


\subsection{NOMA Pairing}
A major challenge of fast grant is the prediction of the active MTDs. Although the prescribed MAB learning approach has a potential to efficiently handle the resource allocation to MTDs with limited information, its performance is highly dependent on the preceding prediction step. Particularly, MAB performance will suffer with the increase of the number of MTDs selected by the BS for fast grant while they are actually inactive. To reduce the resource wastage due to prediction error, and attain better utilization of the limited resources in massive MTC, a 2-user NOMA technique is adopted. 

In the proposed NOMA technique, at each cycle, each granted device using MAB is assumed as a CH that seek pairing with other non-CHs (nCH). In this context, nCHs are all other active MTDs that were not granted an uplink resource. The pairing is initiated by the CHs by sending a pairing request, then receive responses from {\it eligible} nCH(s). The eligibility between CHs and nCHs is specified by two pairing phases. First, the association phase where each nCH associate itself to the nearest CH for a chance to be paired to it. This could be done by measuring the received signal strength (RSS) of the pairing request and select the CH with the highest RSS. Although restrictive, this phase helps to reduce the interference and allow the CHs to lower the power needed for sending the pairing request. It also reduces the communication overhead needed for the distributed pairing by forcing a single eligible CH for each nCH which is more suitable for massive MTC. However, this could be easily relaxed by allowing each nCH to associate itself to multiple CHs according to their RSS. The second phase is the pair selection. This phase depends on a signal-to-noise ratio (SNR) threshold, $\gamma_{th}$, that guarantees minimum distinction between the signals of the paired devices at the BS for smooth SIC operation. In this regard, we assume that the BS announces a tolerance power, $P_{tol}$, as the minimum power difference required for efficient SIC operation. Also, we assume that each MTD is able to acquire the CSI of its link to the BS from the pilot signal sent from the BS.

Based on the above, we have the following proposed pairing scenario. Each CH transmits a pairing request that contains its ID and a SNR threshold, $\gamma_{th}$, to declare itself as a CH. Then, other nCHs respond to their eligible CHs. Considering a 2-user NOMA system, each CH {\it randomly} selects one of its responded nCHs to share the uplink grant with it. The value of $\gamma_{th}$ could be derived from the following formula:
\begin{eqnarray*}
P_s \gamma_s-P_w \gamma_w \ge P_{tol},
\end{eqnarray*}
where $P_s,\gamma_s $ are the transmission power and the normalized channel gain between the strong device and the BS, respectively. Similarly, $P_w, \gamma_w$ are for the weak MTD. Assuming equal power allocation where $P_s = P_w= P_t$, then $\gamma_{th}$ is:
\begin{equation}
\gamma_{th}= 
\begin{cases} 
\gamma_{CH}+\frac{P_{tol}}{P_t}, ~~~~ \text{if }\gamma_{CH}= \gamma_w (weak),\\
\gamma_{CH}- \frac{P_{tol}}{P_t} , ~~~~ \text{if } \gamma_{CH}= \gamma_s (strong).
\end{cases}
\end{equation}
To sum up, an MTD $n$ can be paired to its nearest CH $c$ iff:
\begin{gather}
\gamma_{n}\ge  \gamma_{th}, \text{if }\gamma_{c}= \gamma_w (weak), ~~mode=0,\\
 \gamma_{n}\le \gamma_{th} , \text{if } \gamma_{c}= \gamma_s  (strong), ~~mode =1.
\end{gather}
An extra control bit $mode$ could be added to identify what type of members the CH is seeking. For throughput maximization, we assume that $mode=0$ is the default where the selected CHs always consider themselves as the weak users and seek stronger nodes for NOMA pairing provided that $P_{tol}$ holds.  
 Although $mode=0$ results in enhanced system throughput, the CH is susceptible to SIC error propagation. For a network of both mMTC and ultra reliable low latency communications (URLLC) MTDs, $mode=1$ is recommended to protect the CH selected by the BS that is most probably would be URLLC MTD that needs service priority. In general, if a network operates on certain mode but some CHs did not find eligible pairs, it may be allowed to switch mode to find a pair. 

Under the prescribed scenario, if an inactive CH was granted a resource, it will seek pairing as well. Thus, the resource is not wasted with an incentive of an increase in the CH's cumulative reward. Particularly, with NOMA, the total system reward at each cycle is the sum resulted from both CHs and nCHs which could reach double the achieved rewards without NOMA. However, the following cumulative rewards of each CH $c$ and its nCH pair $n$ are used in \eqref{Psi_eqn}:
\begin{eqnarray}
\label{NOMA_reward_acc_CH}
z_c(t)= & z_c(t-1) + \theta_c (t) + \rho \theta_n(t)  \\
\label{NOMA_reward_acc_nCH}
z_n(t)= &z_n(t-1) + (1-\rho) \theta_n(t),  
\end{eqnarray} 
where $\rho$ is a weight factor defining the share of the nCH's reward that goes to the paired CH as an incentive. For a general MTD $i$, the values of other parameters in \eqref{Psi_eqn} are adjusted independently based on its individual reward $\theta_i$. For instance, $n_i$ is only incremented for non-zero reward MTDs. 
This setting overcomes the prediction inaccuracy as the active nCHs themselves respond to the pairing request.


\section{Optimal NOMA} 
\label{optNOMA}
The regret defined in \eqref{old_regret} gives an indication of how far the system performance is from the optimal one.   
For NOMA, to get a meaningful regret, we need to find the optimal NOMA pairing that maximizes the network reward as a reference performance which is known to be complex for massive MTC \cite{pairing_uplinkNOMA}. However, our objective is to find a reference performance that is comparable to the proposed 2-step scheme  
to specifically help in improving the second step of NOMA pairing while keeping the benefits of the learning step. 
 Hence, we formalize a {\it quasi-optimal} NOMA scenario. In this scenario, the CHs selected at the first step would be inputs to the problem and it is required to find the optimal pairs for the given CHs. The CHs are either the ones with the highest rewards or MAB-selected. This formulation helps to assess the performance of the proposed NOMA with random pairing for future improvement of the 2-step scheme. 

Based on that, the problem is formalized as a binary integer programming (BIP) problem with the objective of maximizing the total system reward:  
\begin{eqnarray}
\label{opt_NOMA}
\max&\hspace{-1cm} \sum_{c=1}^{M} \sum_{n=1}^{N_{an}}  \omega_{c,n} I_{c,n}  \\  
\label{C1}
\text{s.t.} &  \sum_{n=1}^{N_{an}}  I_{c,n} \le 1, ~~~~~~~\forall c   \notag \\ 
\label{C2}  
&  \sum_{c=1}^{M} I_{c,n} \le 1 ,  ~~~~~~~\forall n     \notag \\ 
\label{C3}
&~~~~  I_{c,n} \in \{0,1\},  ~~~~~~~~\forall n, c       \notag  \\  
&  \abs{P_n \gamma_n-P_c \gamma_c}  \ge P_{tol} ,~~~\forall n, c     \notag  
\end{eqnarray}
where $I_{c,n}$ is a pairing binary variable takes the value $1$ when CH $c$ is paired with nCH $n$. The nCH index $n$ is set to span all actual active $N_{an}$ nCHs, whereas a maximum of $M$ CHs, already selected at the first step, exist in the network at each cycle. 
The first two conditions are to assure that each CH is paired to only one nCH and vice versa. The last condition is the pairing condition required for successful SIC decoding, and the absolute value is used to indicate both modes $0,1$. 
Maximizing the rewards is done via the optimization weights $\omega_{c,n}= \theta_c+ \theta_n$. 
To satisfy the last condition, we could set $\omega= 0$ while solving the problem if the condition does not hold for any pair. The problem was solved using Matlab.

\section{Simulations and Discussions}
\label{sim}
In the following, we consider $N$ MTDs randomly located at fixed points of a square area with side length $500$ meters. At the beginning, the activation of the MTDs follows the Beta distribution with parameters ($\alpha=3, \beta=4$) within $I_A=10$ slots.
After the first activation of the $N$ MTDs, we randomly select MTDs for reactivation at each cycle to keep a dynamic activation process. Note that, by considering a case of $N>>M$ the overload situation sustains. For the channel modeling parameters, a noise power is considered to be -$174$ dBm/Hz,
bandwidth is $360$ kHz and standard deviation for the log normal shadow fading is $10$ dB \cite{MAB_GLBCM}. The traffic predictor is assumed to have $e_p \sim \mathcal{N}(0.01, 0.04)$, and  $P_a \in [0.8,1]$.

Regarding the utility function \eqref{utility_fn}, we set $\delta_1=0.2, \delta_2= 0.3, \delta_3= 0.5$, where the delay gets the highest weight. Additionally, we set the parameters of the function in \eqref{Gomp_delay} as $ a= 1 , b= 8, c= 0.03$ \cite{MAB_GLBCM}. To better illustrate the ability of MAB learning in achieving the essential delay requirement of MTC, we chose to set the first $N/2$ MTDs to be with strict maximum delay of $D_i \in [1,100]$ slots, whereas the other $N/2$ MTDs are with relaxed maximum delay of $D_i \in [150,300]$ slots. 
Also, we assumed that the rate threshold is satisfied for all MTDs (i.e. $\mathbbm{1}[r_i(t) \ge R_{i_{min}}] =1$ in \eqref{Rwd_eqn}). Table.~\ref{SMP} shows the parameters used in the simulations.

\begin{table}[!t]
\vspace{3mm}
\caption{System model parameters}     
\label{SMP} 
\centering
\begin{tabular}{|c|l|c|}
\hline
\textbf{Parameter}      &\textbf{Definition}& \textbf{Value}            \\ \hline
$M$                                  &Number of uplink resources (RBs)& 10                \\ \hline
$N$                                 & Number of MTDs & 500       	    \\ \hline
$T$                                 & Total number of cycles & 10000       	    \\ \hline
$P_t$                    & Transmission power&   10 dBm\\ \hline
$P_{tol}$			& Detection threshold for SIC &  4 dBm \\ \hline
$\rho$                    & Weight for nCH reward division&   0.3\\ \hline
\end{tabular}
\end{table}

\begin{figure*}[t]
  \centering

  \subcaptionbox{Number of times each MTD was scheduled in both OMA and NOMA \label{hist_all_both}}[.32\linewidth][c]{%
    \includegraphics[width=62mm, height=43mm]{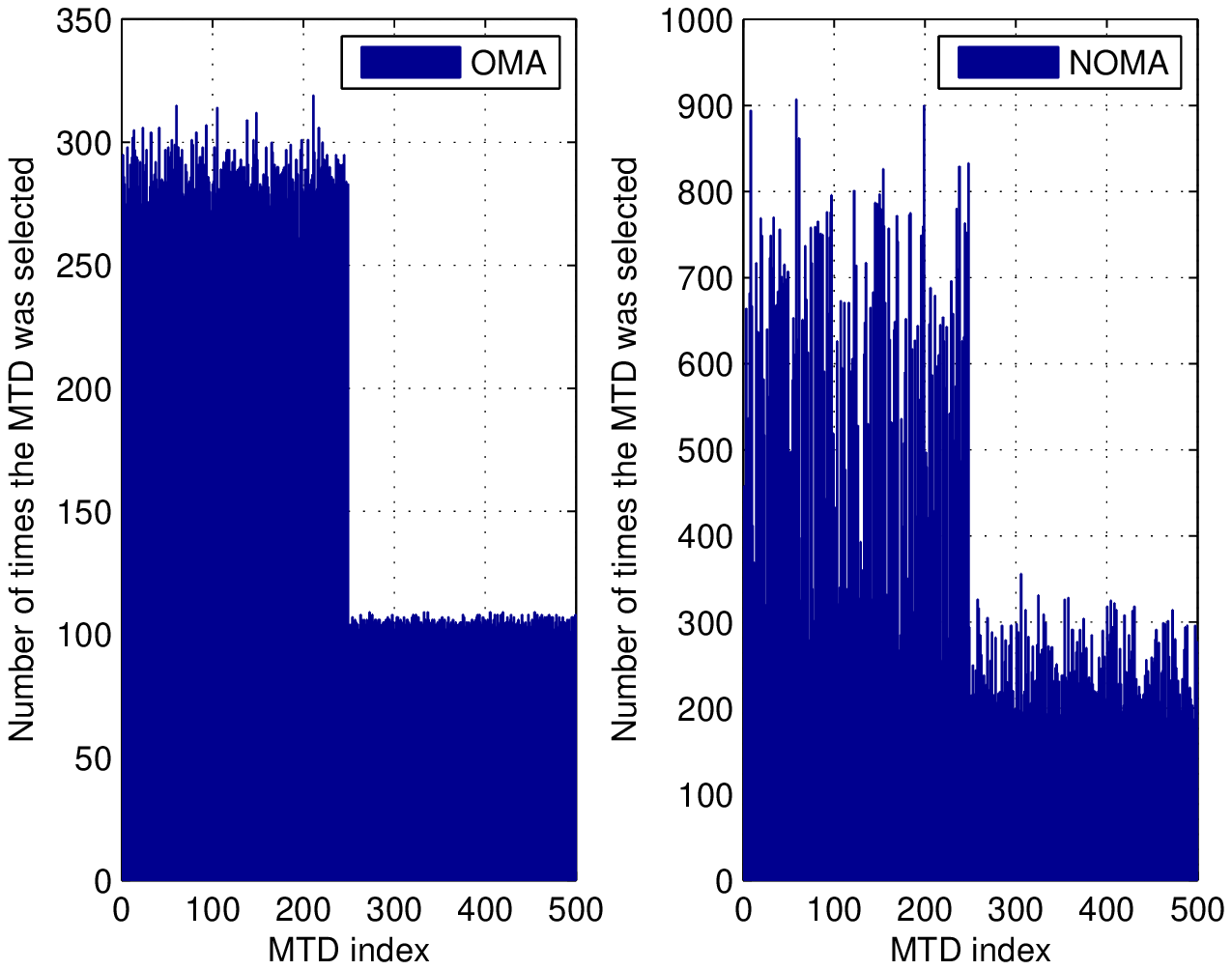}}\quad 
  \subcaptionbox{Total waste of resources \label{missed_rscs}}[.32\linewidth][c]{%
    \includegraphics[width=62mm, height=43mm]{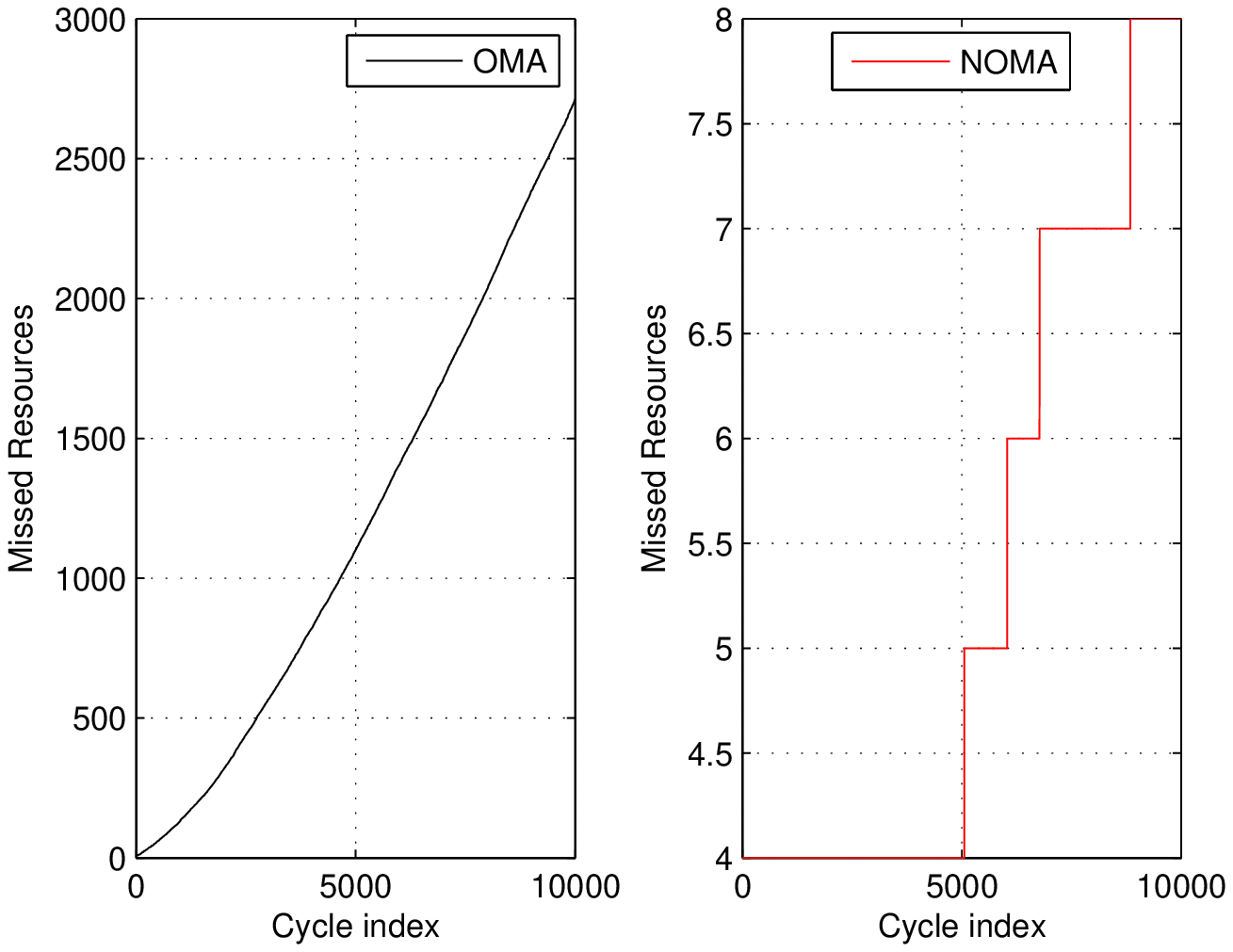}}\quad
\subcaptionbox{Rewards of both systems (b) The difference (Best - MAB) gives the regret. \label{Reward_both}}[.32\linewidth][c]{%
    \includegraphics[width=62mm, height=43mm]{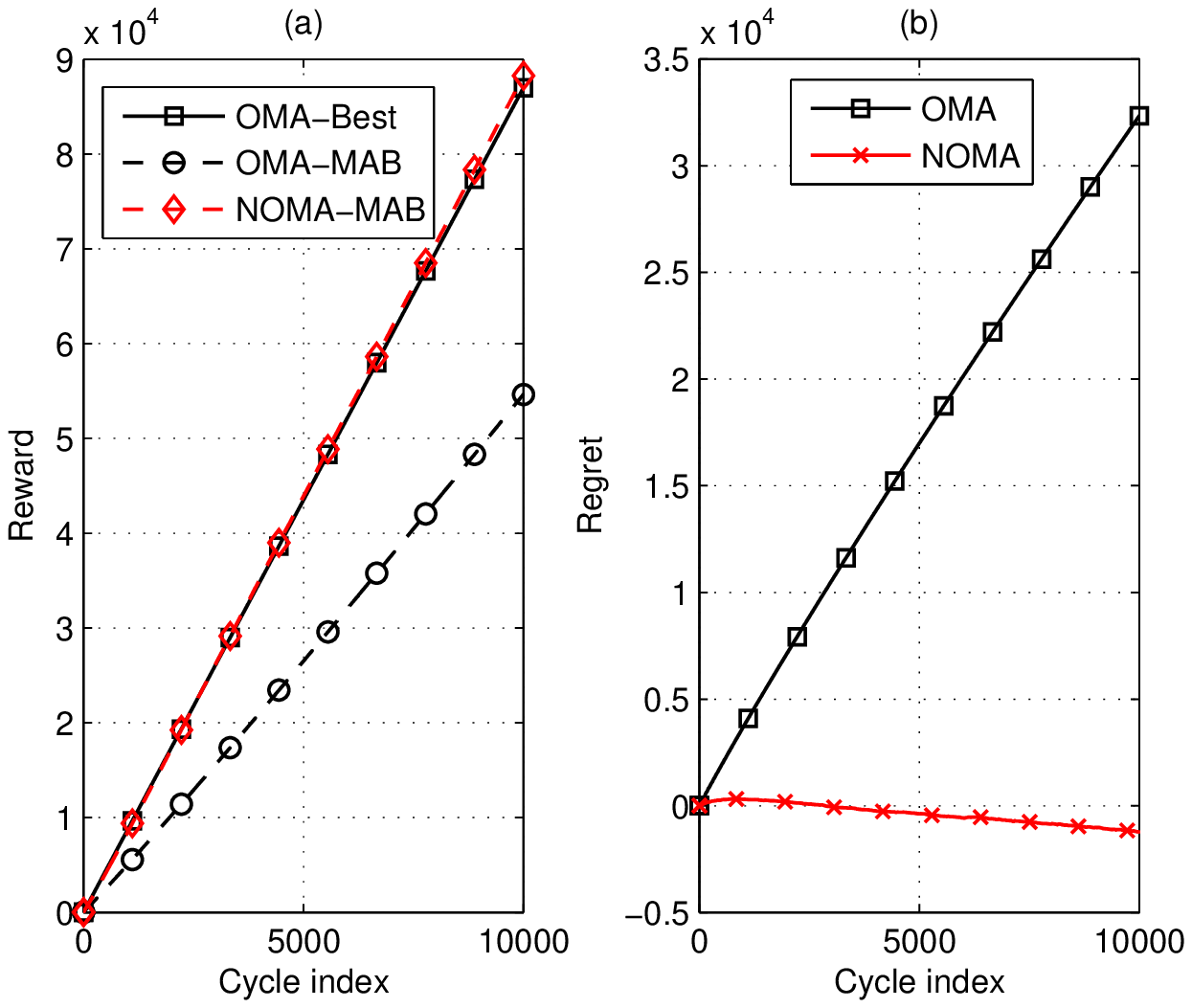}}\quad

  \bigskip

\subcaptionbox{Further enhancements to NOMA system \label{opt_MS}}[.32\linewidth][c]{%
    \includegraphics[width=62mm, height=43mm]{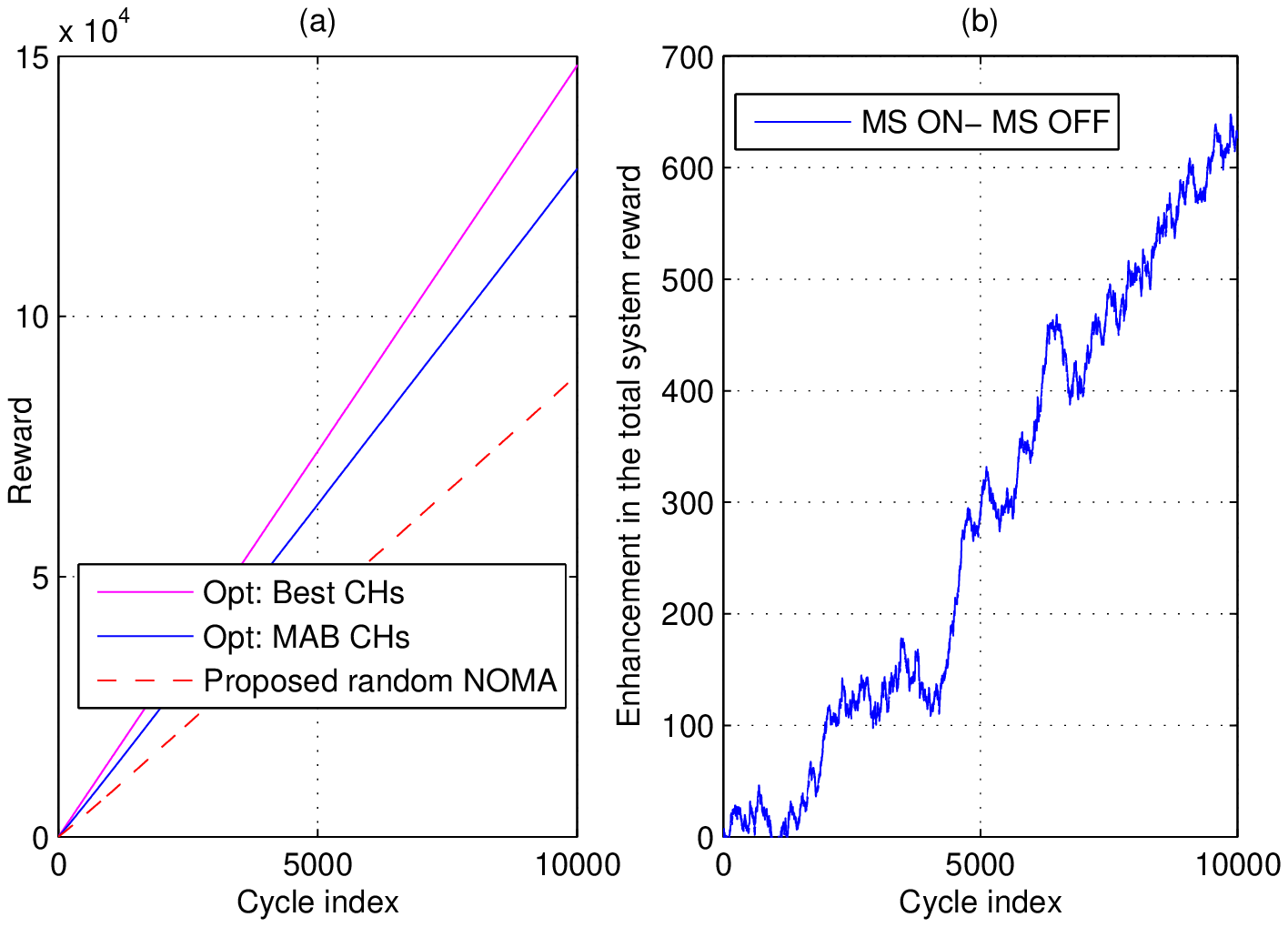}}\quad  
  \subcaptionbox{Effect of the prediction error on the resource wastage \label{missed_rscs_diffPrederr}}[.32\linewidth][c]{%
    \includegraphics[width= 62mm, height=43mm]{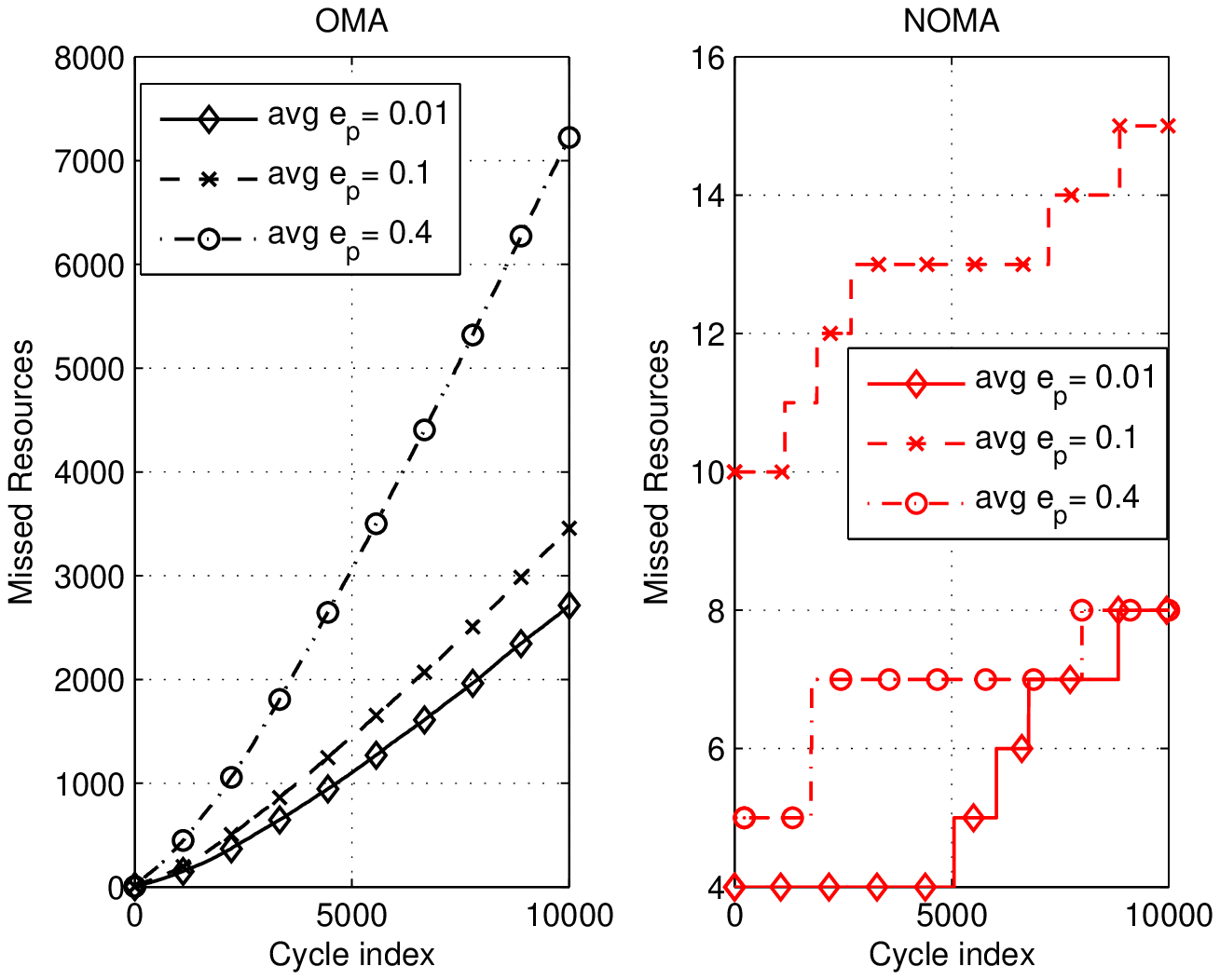}}\quad
\subcaptionbox{Effect of the prediction error on the system reward \label{Rewards_both_diffPrederr}}[.32\linewidth][c]{%
    \includegraphics[width=62mm, height=43mm]{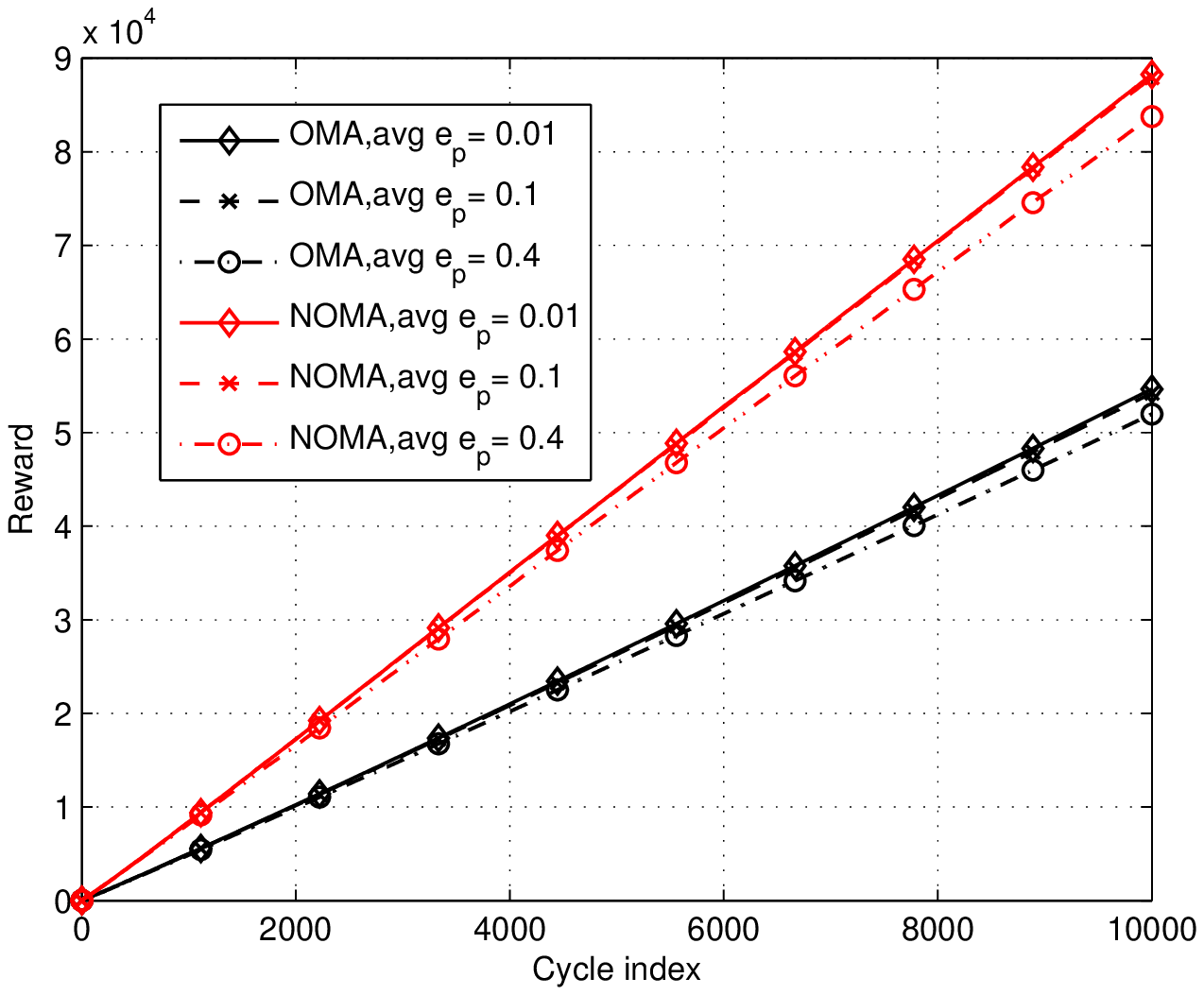}}\quad

\label{all}
\end{figure*}

\subref{hist_all_both} plots the histogram of the number of times each MTD was scheduled for both OMA and NOMA systems. In both cases, it is shown that the MAB learning technique is able to schedule the MTDs with strict delay requirements more frequently (i.e. first $N/2$ MTDs). However, with NOMA, the total number of scheduling times is higher. This is due to the better utilization of the resources offered by NOMA. 
This was also verified in \subref{missed_rscs} where the accumulated number of missed resources with time for both cases are plotted. The figure depicts the enhancement achieved by the proposed NOMA in utilizing the resources missed due to prediction errors. The staircase curve of NOMA case is a result of the infrequent wastage of resources.  

Regarding the rewards and regret, in \subref{Reward_both}, we take the optimal rewards achieved by scheduling the available highest-reward MTDs, one per each RB, as the minuend of \eqref{old_regret}. This is the curve labeled as ``Best'' in \subref{Reward_both}-(a). The subtrahends are the achieved rewards with MAB only and MAB with NOMA for OMA and NOMA regrets, respectively. \subref{Reward_both}-(a) depicts the improvement gained from NOMA that exceeds the best OMA performance. 
This can also be seen in the regret curves shown in \subref{Reward_both}-(b). Although the reference optimal reward here is not related to NOMA, it is good to see how far our practical proposed NOMA-MAB scheme is from the impractical best non-NOMA case that needs the gathering of all MTDs' information and QoS requirements at the BS to be able to optimally allocate the resources.  

Defining the missing ratio as the ratio between the number of missed resources to the total number of available resources during the whole period (i.e. $MT$), Table~\ref{perf_table_t_dEST} compares the performance of OMA and NOMA. The table depicts that NOMA offers better utilization of the resources where the missing ratio significantly decreases. It also shows the achieved increase in the number of served MTDs (i.e. winners) along with an enhancement in their average access delay. The uniform selection of nCHs pairs results in a slight increase in the average maximum tolerable delay of the scheduled MTDs.

\begin{table}[!t]
\vspace{3mm}
\caption{Performance Evaluation with $t^\prime$ estimated}     
\label{perf_table_t_dEST} 
\centering
\tiny
\begin{tabular}{|l||c|c|c|c|}
\hline
\textbf{System} & \textbf{Missing Ratio} & \textbf{Winners}& \textbf{Avg. Max Delay} &     \textbf{Avg. Access Delay}           \\ \hline\hline
  OMA                     &   $0.0271$  & $9.7288\times 10^4$ &   $105.1120 $ &  $36.6666 $ \\ \hline   
   NOMA                     &     $0.0001$ &  $19.0483\times 10^4$ &  $108.5102$ &$19.0421$\\ \hline 
\end{tabular}
\end{table}

Focusing on NOMA, \subref{opt_MS}-(a) compares the reward resulted from the proposed NOMA with simple random pairing to the rewards resulted from the 2-step optimal NOMA (i.e. quasi optimal) pairing analyzed in section~\ref{optNOMA}. We consider two inputs for optimal pairing, the first is the best CHs with the highest rewards, and the second is the MAB-selected CHs. These input CHs will seek pairing using \eqref{opt_NOMA}. The gaps between each of these two curves and the proposed NOMA represent the further enhancement that can be achieved by combining NOMA with fast grant in a 2-step fashion as proposed in this manuscript. Specifically, the gap between the proposed NOMA curve and the optimal with MAB CHs indicates the further enhancement that could be achieved by enhancing the pairing scheme. 
 On the other hand, to enhance the performance of random NOMA, we added a mode switch (MS) function to the system. This function allows each CH to switch its pairing mode temporarily if it failed to find a NOMA pair with the nominal mode of the network. Specifically, for a network operating at $mode=0$, each CH that does not find a pair at certain cycle is allowed to re-announce itself as a CH operating at $mode=1$ at this specific cycle. This is to increase the resources utilization and not to miss the pairing opportunity. \subref{opt_MS}-(b) depicts the difference in the system reward while the function is ON and OFF. However, this enhancement results in an increase of the pairing overhead and time. 
The mode switch function could be useful for networks with strict pairing conditions or non-dense devices where the probability of not finding a pair increases.

Using different average prediction errors $\bar{e}_p= 0.01, 0.1, 0.4$, we examine the effect of the predictor efficiency on the OMA and NOMA.  
\subref{missed_rscs_diffPrederr} shows a relatively large increase in the number of wasted resources of OMA system with the increase of $\bar{e}_p$ which is not the case for NOMA. This reflects that NOMA made the system less vulnerable to the predictor efficiency. For the rewards illustrated in  \subref{Rewards_both_diffPrederr}, although NOMA reward is always higher than OMA with different $\bar{e}_p$ values, increasing $\bar{e}_p$ degrades the performance of both systems. For OMA, the performance degradation is due to the loss of the resources. In contrast, with NOMA the reason is the random selection of the nCHs which represent larger portion of the selected MTDs with high $\bar{e}_p$. This can be improved by enhancing the pairing scheme in our future work. Overall, the results show a potential of performance improvement at affordable complexity.

\section{Conclusion}
\label{conc}
We introduced a communication framework to provide access to MTC devices based on fast uplink grant with NOMA. The proposed 2-step scheme, employs MAB learning technique at the first step to schedule the devices based on different QoS requirements including latency. Then, it allows the selected devices to randomly pick a NOMA pair from an eligible set of active devices in a distributed manner. This technique decouples the scheduling efficiency and the predictor performance. We build a simulation model for a Beta-distributed traffic whereas the traffic predictor was abstracted. The simulation results show the effectiveness of the proposed technique in providing access to more MTDs in a timely manner while accommodating their heterogeneous QoS requirements. Additionally, the results illustrate significant enhancement in the scarce resources utilization. It was also depicted that the proposed scheme is more robust against source traffic prediction errors. A quasi-optimal 2-step NOMA formulation was provided as a benchmark for future performance enhancement.



%

\end{document}